# Spin-Orbit Torque Flash Analog-to-Digital Converter

Hamdam Ghanatian[1*], Luana Benetti[2], Pedro Anacleto[2], Tim Böhnert[2], Hooman Farkhani[1], Ricardo Ferreira[2], Farshad Moradi[1]

Although Analog-to-digital converters (ADCs) are critical components in mixed-signal integrated circuits (IC), their performance has not been improved significantly over the last decade. To achieve a radical improvement (compact, low power and reliable ADCs), spintronics can be considered as a proper candidate due to its compatibility with CMOS and wide applications in storage, neuromorphic computing, and so on. In this paper, a proof-of-concept of a 3-bit spin-CMOS Flash ADC using in-plane-anisotropy magnetic tunnel junctions (i-MTJs) with spin-orbit torque (SOT) switching mechanism is designed, fabricated and characterized. The proposed ADC replaces the current mirrors and power-hungry comparators in the conventional Flash ADC with seven parallel i-MTJs with different heavy metal (HM) widths. Monte-Carlo simulations based on the experimental measurements show the process variations/mismatch limits the accuracy of the proposed ADC to 2 bits. Moreover, the maximum differential nonlinearity (DNL) and integral nonlinearity (INL) are 0.739 LSB (least significant bit) and 0.7319 LSB, respectively.

ADCs translate analog input to digital output, and play a crucial role in computational systems[1-4]. With emerging computing in memory (CiM) for implementation of deep neural networks (DNN), the need for compact and low-power ADCs is increasing[5-7]. The conventional ADCs suffer from technology scaling due to the large process variation and lower performance in scaled nodes. According to the recent published roadmap for ADC, the ADC performance shows no obvious improvement in terms of resolution, area, and power consumption in the next few years using the current technology[8]. One promising solution can be shifting from conventional complementary metal-oxide-semiconductor (CMOS) technology to new hybrid technologies such as spin-CMOS technology[9].

Magnetic tunnel junction (MTJ) is a promising candidate as a spintronic device for many applications due to its compatibility with CMOS, non-volatility, high retention time and long endurance[10,11,12]. An MTJ consists of an oxide layer sandwiched between two ferromagnetic (FM) layers. The magnetization direction of one of the FMs is fixed and it is called pinned layer (PL) while the other one that can be switched along its easy axis is called the free layer (FL). If the magnetization directions of the FL and PL are parallel, the device is in parallel state (P-state)-state, where the MTJ presents a low resistance (logic '0'), whereas, if the magnetization direction of the FL is in the opposite direction of the PL, the device is in antiparallel state (AP-state) and shows a high resistance (logic '1'). The magnetic orientation of the FL can be adjusted by passing a charge current ($I_{STT}$) through the MTJ via spin-transfer torque (STT) mechanism[13]. However, one of the challenges with this method for switching is that the thin oxide layer can be broken when the device experiences a high amount of $I_{STT}$ leading to the reduction of reliability and endurance of MTJs[14]. Spin-orbit torque (SOT)-based MTJs have been proposed to overcome this issue while improving the switching efficiency[15]. In SOTs, a charge current ($I_{SOT}$) greater than the critical charge current ($I_{SOT,crit}$) flows through a heavy metal (HM) and the switching is accomplished by spin-orbit torque (SOT) through the spin Hall effect (SHE) [16,17].

Recently, several works on designing ADC using SOT-based MTJ have been reported[8,18-21]. Jiang et al.[8] have developed a spintronic ADC based on SHE and voltage-controlled magnetic anisotropy (VCMA). To tune $I_{SOT,crit}$ of each MTJ, a resistive ladder is utilized to provide different voltages on the MTJs. Such approach suffers from power overhead and reliability issues[18]. In others works[18-21], a taper HM is shared between MTJs in which the width of the HM ($w_{HM}$) is engineered to tune $I_{SOT,crit}$. To sense the state of each MTJ in such approaches, a current flows through the MTJ ($I_{Sens}$). However, considering the fact that the shared HM forms the bottom contact of the MTJs, $I_{Sens}$ will pass through only a part of the HM. MTJs will experience different bottom contact resistance depending on their position on the shared HM. It is worth noting that different HM widths, obviously, leads to different HMs resistances in the path and this resistance gets larger for MTJs placed far from the HM terminal connected to the ground. The larger the resistance of the HM in the current path, the larger the degradation of the magnetoresistance (MR) and thereby reading reliability. To overcome this issue, some works use side-reading approach [18-19], while other works use a dummy

[1]Department of Engineering Aarhus University, 8200 Aarhus, Denmark. [2]International Iberian Nanotechnology Laboratory (INL), Av. Mestre Jose Veiga s/n, Braga 4715-330, Portugal.

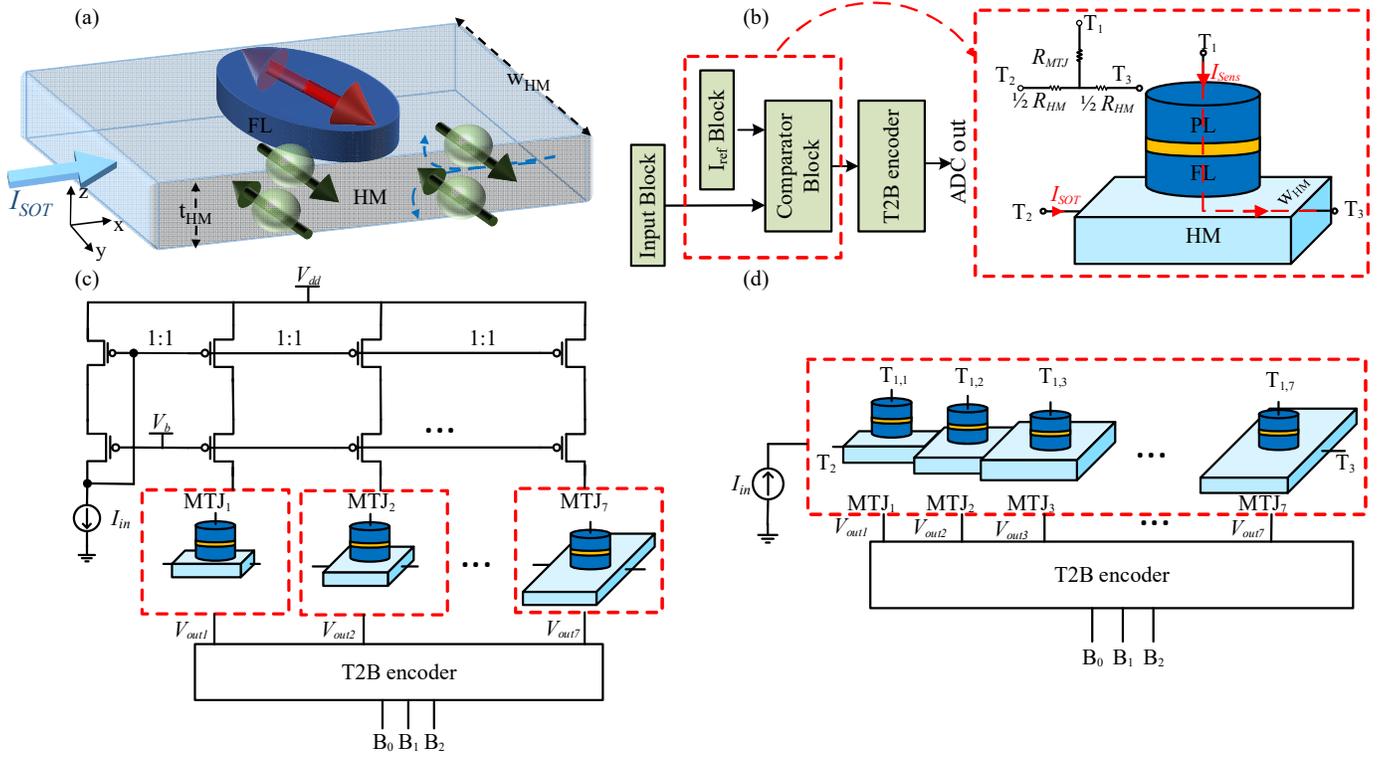

Fig. 1 (a) The concept of SOT switching (b) The block diagram of the current-mode Flash ADC. The $I_{ref}$ and comparator blocks can be replaced with SOT-based MTJ. (c) 3-bit spin-CMOS Flash ADC (parallel design) (d) 3-bit spin-CMOS Flash ADC (serial design).

quantizer to sense each MTJ resistance [20]. The difference in resistances of the adjacent HMs is compensated by adjusting the size of the transistor in the sensing circuit[21]. However, in the proposed solutions, increasing the complexity of the sensing circuit is the cost of mitigation issue of MR degradation.

In this paper, the proof-of-the-concept of implementing an ADC based on spintronic devices is investigated which provides design guidelines for future spin-CMOS ADCs. To this end, a spin-CMOS ADC is proposed, designed, and characterized in which SOT-based MTJ and its $I_{SOT,crit}$ act as a comparator and reference current ($I_{ref}$) in conventional current-mode Flash ADCs, respectively. In spite of the proposed structures in literatures [18-20], in this structure, in-plane anisotropy MTJs (i-MTJ)s are placed in parallel branches to mitigate the MR deduction and the complexity of the sensing circuit. The impact of the HM resistance on the MR is shown by comparing the measurement data extracted from the structure proposed in literature[19] with the approach presented in this paper. The measurement results show that the MR values of the proposed ADC are more than those of the structure in[19] which means the reading reliability can be improved in the proposed structure. The input current ($I_{in}$) is copied to each branch and in case $I_{in}$ is higher than $I_{SOT,crit}$ of the MTJ, the MTJ will switch. Hence, $I_{SOT,crit}$ of each i-MTJ can behave like $I_{ref}$ in the current-mode CMOS Flash ADCs. All i-MTJs are set in the P-state and if $I_{in} > I_{SOT,crit}$, the i-MTJ is switched to the AP-state. The width of the HM ($w_{HM}$) is tuned so that the $I_{SOT,crit}$ of each MTJ is compatible with reference currents ($I_{ref}$, $2I_{ref}$, $3I_{ref}$, ...) of the current-mode CMOS Flash ADC. Furthermore, Monte-Carlo simulation is performed to analyze the impact of the process variations/mismatch of the i-MTJs and transistors on the reference currents of ADC. To this end, a random variable with a Gaussian distribution for each i-MTJ is considered. The mean and standard deviation ($\sigma$) of the variable are defined by the measurement data of the i-MTJs. Moreover, the variations of the CMOS circuit (the current mirror of $I_{in}$) has been included to extract the $I_{SOT,crit}$ values.

## Spin-CMOS ADC

The principle of the SOT switching mechanism in the FL of the SOT-based MTJ is shown in Fig.1 (a). In this structure, a charge current ($I_{SOT}$) flows through the HM along the x-direction. The SHE in the HM creates a pure spin current in z-direction, which is magnetized along the y-direction. This pure spin current generates an STT, which can switch the FL magnetization at a critical spin current density ($J_{s,crit}$), which is similar for all MTJs that are nominally identical. The conversion efficiency between the charge current density and the spin current density is described by the spin Hall angle $\theta$. So, the $I_{SOT,crit}$ can be described by[22-24]

$$I_{SOT,crit} = J_{SOT,crit} t_{HM} w_{HM} = \frac{2e}{\hbar} \frac{J_{s,crit}}{\theta} t_{HM} w_{HM} \quad (1)$$

with the critical change current density ($J_{SOT,crit}$), the electrons charge e, the electrons spin expressed by the reduced Planck's constant $\frac{\hbar}{2}$ and the HM thickness $t_{HM}$. Thus, the charge current required for switching is proportional to $w_{HM}$, which makes tuning of the critical charge currents relatively easy in these devices.

The schematic of the current-mode Flash ADC which consists of the input, $I_{ref}$, comparator, and thermometer code to binary (T2B) encoder blocks are depicted in Fig.1 (b). Flash ADCs are categorized into two groups: 1) voltage mode and 2) current mode.

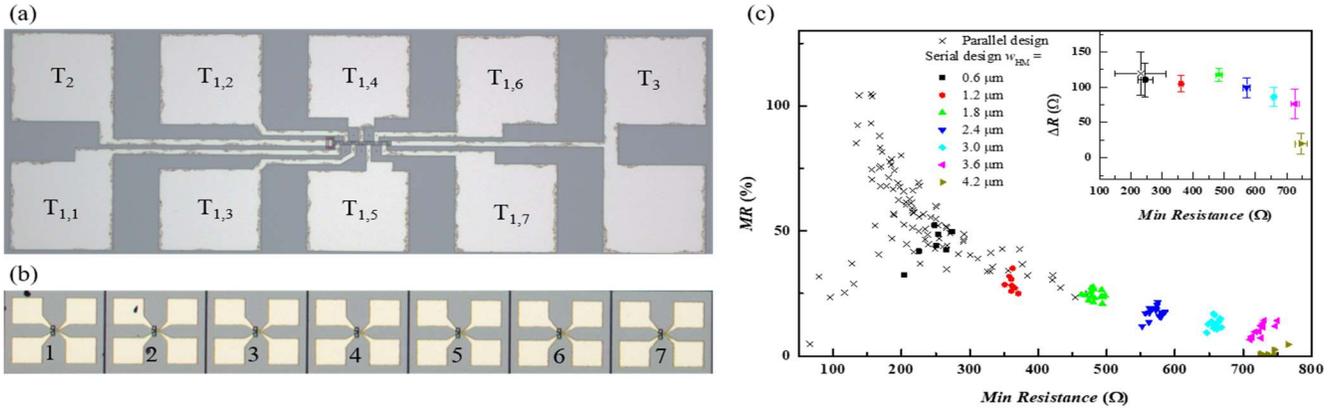

Fig. 2 (a) Images from optical microscope of the serial design and (b) parallel design. (c) MR as function of the minimum resistance for serial and parallel designs for different $w_{HM}$, inset the resistance variation.

Current-mode Flash ADCs have some advantages over voltage-mode ADCs, such as less power consumption and the ability to operate with smaller supply voltages[21]. The input block makes several copies from $I_{in}$, then the comparator block compares these copies with reference currents coming from $I_{ref}$ block. The outputs of the comparator block are encoded by the T2B encoder and binary data corresponding to the input signal is generated as the ADC output. Hence, in the n-bit current-mode CMOS Flash ADC, $2^n-1$ copies of $I_{ref}$ with different weights (i.e., $I_{ref0}$, $2I_{ref0}$, …, $(2^n-1)I_{ref0}$) and $I_{in}$ are required. The main idea of the proposed work is to replace the current mirror circuits needed for generating different copies of $I_{ref}$ as well as the comparator block by an SOT-based MTJ as shown in Fig.1 (b). Since $I_{ref}$ values are multiplications of $I_{ref0}$, the size of transistors in the current mirror circuit will progressively increase. By replacing $I_{ref}$ and comparator blocks with an SOT-based MTJ, space and mismatch issues can be mitigated. As shown in Fig.1 (b), $I_{SOT}$ as an input current ($I_{in}$) flows through the HM from $T_2$ to $T_3$ and as mentioned before the SOT-based MTJ acts as a comparator; hence it compares the $I_{in}$ with its $I_{SOT,crit}$ (behaves as the $I_{ref}$ block). To sense the MTJ resistance, a current ($I_{Sens}$) passes through the MTJ and a part of the HM from $T_1$ to ($T_2/T_3$). The 3-bit spin-CMOS Flash ADC in two different designs called parallel and serial designs are shown in Fig.1 (c) and (d), respectively. In both, seven SOT-based MTJs are utilized to create an ADC with 3 bits of resolution. By engineering the $w_{HM}$, $I_{SOT,crit}$s can be tuned so that by increasing $w_{HM}$, the required current for switching the SOT-based MTJ will increase[25]. To this end, $w_{HM}$ of each MTJ should be properly designed to ensure that $I_{SOT,crit}$s for SOT-based MTJ$_1$, SOT-based MTJ$_2$, …, SOT-based MTJ$_7$ are equal with $I_{SOT,crit}$, $2I_{SOT,crit}$, $3I_{SOT,crit}$, …, and $7I_{SOT,crit}$, respectively. In the serial design[18-20], the SOT-based MTJs are put in series through HMs. As shown in Fig. 1(d), by using this design, the input block (shown in Fig. 1(b)) that consists of the $I_{in}$ mirror branches can be removed. However, the HM resistance (depending on the MTJ position) degrades the MR and the reading reliability. For instance, if $T_2$ (Fig. 1(d)) is connected to the ground, the sensed resistance by $I_{Sens}$ from $T_{1,7}$ to $T_2$ according to the equivalent resistive network of the SOT-based MTJ depicted in Fig. 1(b) that is $R_{MTJ7} + 1/2\, R_{HM7} + R_{HM6} + … + R_{HM1}$. Therefore, the MR for SOT-based MTJ$_1$ is $R_{MTJ7}(AP)-R_{MTJ7}(P))/(R_{MTJ7}(P) + 1/2 R_{HM7} + R_{HM6} + … + R_{HM1})$ where, $R_{MTJ}(AP)$ and $R_{MTJ}(P)$ are the SOT-based MTJ resistance when SOT-based MTJ is in AP-state and P-state, respectively. Moreover, the different resistance seen from $T_1$ of each MTJ leads to an increase in the complexity of the sensing circuit. To mitigate this issue, a parallel design, as shown in Fig. 1(c), is proposed in this paper. In this structure, SOT-based MTJs are detached and the HM resistance seen from $T_1$ of each MTJ is almost equal if all MTJs are in the same states. However, $I_{in}$ should be copied by current mirrors (the input block) and fed into each of the SOT-based MTJs. In both designs, the result of the comparison between $I_{in}$ and $I_{SOT,crit}$ in each SOT-based MTJ is presented as a voltage signal ($V_{outi}$ (1 ≤ i ≤ 7)). The T2B encoder block creates a 3-bit digital output ($B_0$, $B_1$, $B_2$) based on $V_{outi}$. The detail of circuit design for sensing of SOT-based MTJ states and T2B are presented in[21].

## Results and discussion

The microscopic images of the serial and parallel designs are shown in Fig. 2 (a) and (b), respectively. Fig. 2(c) shows the MR versus minimum resistance (the resistance seen by $I_{Sens}$ when the MTJ is in the P-state) for the two designs. In the serial design, $T_2$ is connected to the ground. MR dependency with the position of the i-MTJ is observed for the serial design in which the MR difference between the lowest (belongs to MTJ$_7$) and highest (for MTJ$_1$) is around 47%. The MR for the MTJs with the width of 4.2 μm is the lowest as compared to the other MTJs because as mentioned before, the resistance seen from $T_{1,7}$ to $T_2$ is larger. In general, MR in the serial design is lower than that in the parallel design because of the large HM resistance. Moreover, the dependency of MR to i-MTJ position is much smaller in the parallel design because the resistance seen from $T_1$ of each MTJ to the ground is $R_{MTJ}+R_{HM}/2$.

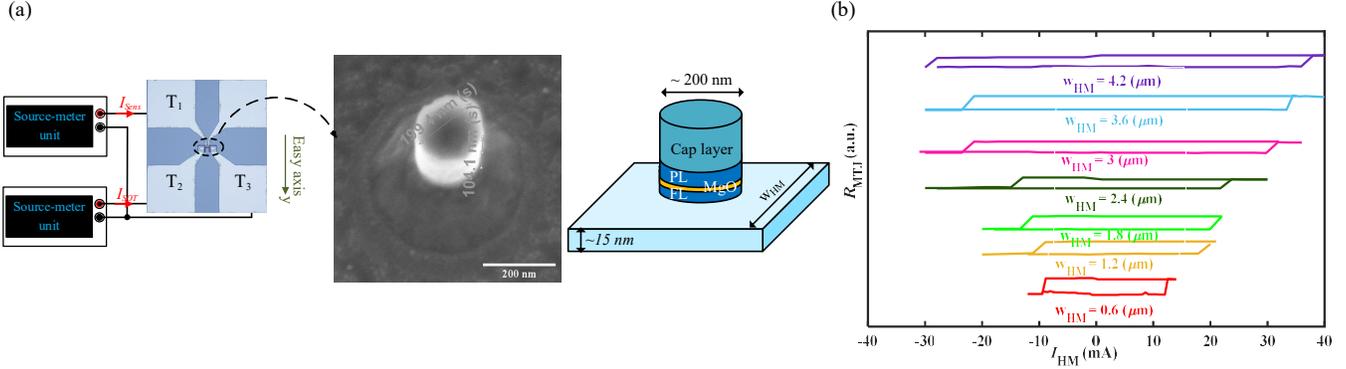

Fig. 3 (a) The schematic view of the experimental setup used for characterization of the SOT- (b) The R-I loop for different $w_{HM}$.

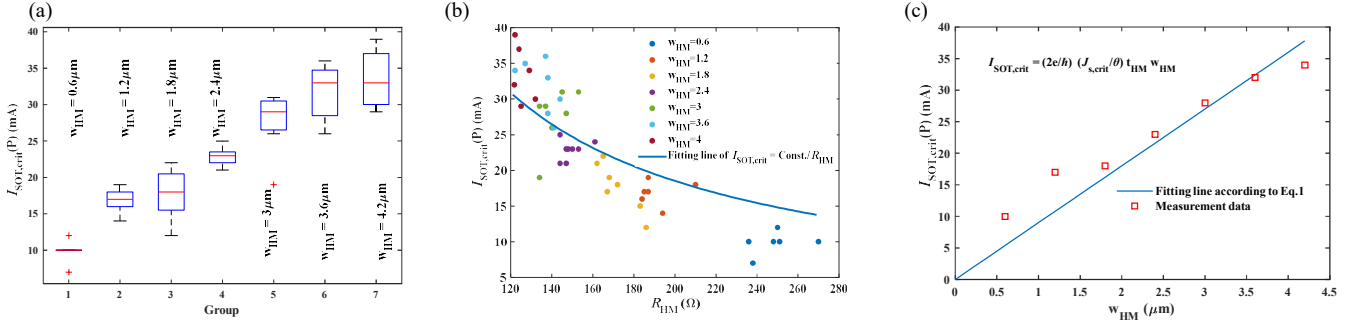

Fig. 4 (a) The box plots of $I_{SOT,crit}$(P) for different $w_{HM}$ (0.6 μm, 1.2 μm, 1.8 μm, 2.4 μm, 3 μm, 3.6 μm, 4.2 μm). (b) The distribution of $I_{SOT,crit}$(P) and $R_{HM}$ for 7 categories of $w_{HM}$. (c) The average of $I_{SOT,crit}$ (P) for each category of $w_{HM}$ versus the nominal value of $w_{HM}$.

The proof of concept of the implementation of a 3-bit Flash ADC based on the spintronic device can be investigated using the measured data from the characterization of the parallel configuration. To this end, the experimental setup of Fig. 3 (a) is utilized to characterize the MTJs. All MTJs are initially set to the AP state by applying an external DC magnetic field with an amplitude of 19 mT along +y. Afterwards, the external magnetic field is removed and $I_{SOT}$ is injected into the HM through $T_2$. Subsequently, $I_{Sens}$ (a DC current) with an amplitude of 100 μA is applied by a source-meter unit to measure the resistance between $T_1$ and $T_3$. This resistance, according to the equivalent resistive network of the SOT-based MTJ (Fig. 1 b) is $R_{MTJ} + 1/2\ R_{HM}$. In this measurement, the samples have been reported that the amount of change in their resistance after switching ($R_{MTJ}$(AP) - $R_{MTJ}$(P)) and their MR are more than 68 Ω and 20%, respectively. Fig. 3 (b) depicts the MTJ resistance versus $I_{SOT}$ in absence of the external magnetic field for 7 SOT-based MTJs with different $w_{HM}$. The positive (negative) current drives switching from P-state to AP-state (AP-state to P-state). In this paper, P-state is considered as the initial state of the MTJ 3-bit spin-CMOS Flash ADC and the switching from P-state to AP-state occurs (during the conversion phase in the ADC [20]) at the critical charge current called $I_{SOT,crit}$ (P). During the reset phase in the ADC, SOT-based MTJs are switched back to their initial states at the critical charge current called $I_{SOT,crit}$ (AP), where the current direction is opposite of $I_{SOT,crit}$ (P). Moreover, as shown in the obtained R-I loops, the width of the R-I loop becomes larger by increasing the $w_{HM}$, which means that, as mentioned in Eq. 1, by increasing $w_{HM}$, the $I_{SOT,crit}$ (AP) and $I_{SOT,crit}$ (P) are rising.

The box plots of $I_{SOT,crit}$ (P) for seven categories are presented in Fig. 4 (a). Each category represents SOT-MTJs with the same size of $w_{HM}$, in which $w_{HM}$ of category 1, 2, …, and 7 is 0.6 μm, 1.2 μm, …, and 4.2 μm, respectively. As shown in this figure, increasing $w_{HM}$ results in an increasing trend in $I_{SOT,crit}$ (P). σ of $I_{SOT,crit}$ for $MTJ_1$, $MTJ_2$, …, $MTJ_7$ is 1.6 mA, 1.7 mA, 3.45 mA, 1.36 mA, 4.16 mA, 3.77 mA, 3.94 mA, respectively. The distribution of $I_{SOT,crit}$ (P) and HM resistance ($R_{HM}$), which are subdivided by seven groups of different $w_{HM}$, are depicted in Fig. 4(b). The trend of increasing $I_{SOT,crit}$ with $R_{HM}$ according to the equation of $I_{SOT,crit}$ (P) = const./ $R_{HM}$ (Eq.1 and $R_{HM}$ = const. / ($t_{HM}$ × $w_{HM}$)) can be observed in this figure. As shown in Fig. 4(a) and (b), categories 1, 2 and 4 ($w_{HMS}$ are 0.6 μm, 1.2 μm and 2.4 μm) have the least variations while the variations of categories 3, 5, 6, and 7 are large and the distributions of these groups overlap with each other and other categories. Such large variations lead to nonlinearity, missing code and low accuracy issues in the ADC design based on the MTJs. Such random distributions are attributed to the variations in the $w_{HM}$, $t_{HM}$ and MTJs. In particular, $t_{HM}$ is thin and the absolute variation is large that results in a large variation of the actual HM current density. Other way around, considering the nominal HM thickness this error results in a variation of the spin Hall angle. The MTJ variation is mainly due to the small size of the nano-pillars with dimensions similar to the grain sizes, which adds a random distribution. $I_{SOT,crit}$(P) versus $w_{HM}$ is presented in Fig. 4(c) in which the square points and the solid line are the measurement data and a fitting line, respectively. In this figure, each point is the average data of each category of $w_{HM}$ that is extracted from Fig. 4(a). The fitting line to the data with 0.8243 of R-squared ($R^2$), represents a linear relation between $I_{SOT,crit}$ and $w_{HM}$ that is mentioned in Eq. 1. This linear dependency enables the linear ADC behavior. From the fitting line, we can determine

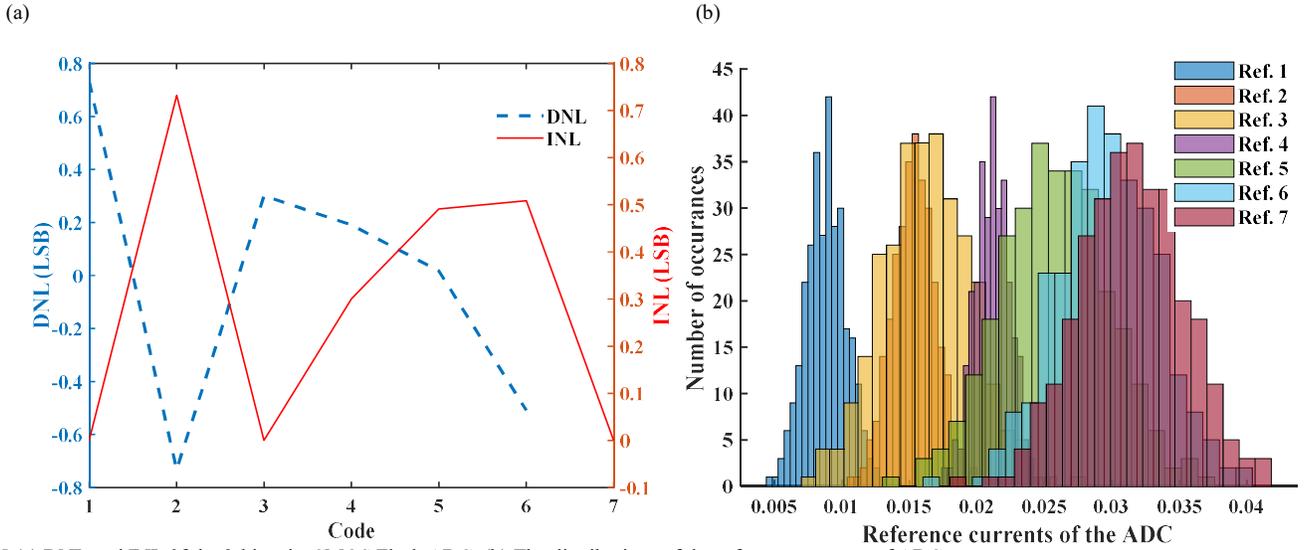
Fig. 5 (a) DNL and INL 0f the 3-bit spin-CMOS Flash ADC. (b) The distributions of the reference currents of ADC.

the characteristic critical current density of the device $J_{SOT,crit} = 0.6 \times 10^{12} \frac{A}{m^2}$, which describes how efficient the SOT current can switch the MTJs, which influences the precision of this ADC.

The differential nonlinearity (DNL) and integral nonlinearity (INL) characteristics for the proposed ADC are shown in Fig.5 (a). The maximum DNL and INL are 0.739 LSB (5 mA) and 0.7319 LSB, respectively. The simulation results are obtained by a behavioral model for MTJs in Verilog-A that is extracted from the measurement. In this model, $I_{SOT,crit}$ for each SOT-based MTJ is the mean value of each category that is extracted from Fig. 4 (c). The CMOS circuits (the current mirrors for $I_{in}$) are simulated using Cadence in TSMC 180nm technology. Monte-Carlo simulation is performed to evaluate the effects of the process variations/mismatch of the MTJs and CMOS circuits on the reference currents of ADC. The distributions of the reference currents shown in Fig. 5(b) are achieved by 300 simulation runs. Each plot includes the distributions of process variations and mismatch of the CMOS circuit of the $I_{in}$ current mirror (Fig. 1(c)) and process variations of the related MTJ. For each MTJ, a behavioral model is considered that contains a variable with a Gaussian distribution. The values of mean and σ of the variable are extracted from Fig. 4 (a). ±2σ yield can be supported only if $MTJ_1$, $MTJ_2$, $MTJ_4$ and $MTJ_7$ are employed while histograms of $MTJ_3$, $MTJ_5$ and $MTJ_6$ strongly overlap with other $I_{SOT,crit}$ distributions. Therefore, according to Fig. 4(b), the maximum available accuracy of the proposed ADC by such fabricated MTJs is 2 bits. The σ for $MTJ_1$, $MTJ_2$, …, $MTJ_7$ are 1.5 mA, 1.6 mA, 3.3 mA, 1.3 mA, 4 mA, 3.7 mA, 3.8 mA, respectively. The values of σ are almost the same ones extracted from Fig. 4 (a) which means the process variation of MTJs is dominant as compared to the process variation and mismatch of the transistors.

## Conclusion

In this paper, SOT-based MTJs are designed, fabricated, and characterized for the implementation of a 3-bit spin-CMOS Flash ADC. The linear relation between $I_{SOT,crit}$ and the width of HM was verified and the figure of merit of the SOT-based MTJ ($J_{SOT,crit}$) is $0.6 \times 10^{12}$ Am$^{-2}$. Seven separated SOT-based MTJs with different width of HMs are employed. In this structure, MTJ and its $I_{SOT,crit}$ play the role of the comparators and $I_{ref}$ blocks in Flash ADC, respectively. Hence, the power-hungry comparators and the current mirrors that generate $I_{ref}$s in current-mode Flash CMOS ADCs are eliminated. The current used for sensing the MTJ resistance, senses the HM resistance of only one MTJ in the path leading to significant improvement in MR and reading reliability. The maximum INL and DNL are in the range of 0.7319 LSB and 0.739 LSB, respectively. Furthermore, Monte-Carlo simulations are conducted for estimation of the ADC accuracy in the presence of the process variation/mismatch of the MTJ and CMOS transistors. The simulation results show the accuracy of the proposed ADC limits to 2 bits, which can be enhanced by improving the MTJ fabrication process in future.

## Methods

An inverted MTJ stack with a 3-terminal geometry, similar to those used in previous works[22,26-27], was proposed. The MTJ consists in 15 W/ 1.4 $CoFe_{40}B_{20}$/ MgO/ 2.2 $CoFe_{40}B_{20}$/ 0.85 Ru/ 2.5 $CoFe_{30}$/ 6 IrMn/ 5 Ru/ 140 Cu/ 30 Ru (thicknesses in nanometer) deposited on Si (100)/ 200nm thermal $SiO_2$ by magnetron sputtering. The MgO thickness was targeted to have a resistance-area product (R×A) of 12 Ω.µm$^2$, since that below 10 Ω.µm$^2$ a tunnel magnetoresistance (TMR) decrease is observed[28]. Through current-in-plane transport measurements, the stack exhibited an R×A of 14.3 Ω.µm$^2$ and 144% TMR. The tungsten in the stack was chosen as heavy metal due to the high spin hall angle reported in the β-phase[29]. However, this phase is only possible for W thicknesses of few nanometers (< 6 nm)[30] which is rather challenging for device fabrication since it reduces the stopping point margin for the pillar etch. By tuning the deposition conditions or incorporating some defects, it is possible to increase the thickness of the β-W[31-33]. In our samples the β-W is limited to 5 nm and we decided to compromise the spin hall angle efficiency in order to have a measurable device.

The nanofabrication process is the same described by Tarequzzaman *et al.*[26]. The electron beam lithography (EBL) was used to pattern 200 nm diameter nanopillars and an ion beam milling system was used for etching. Through the secondary ion mass spectrometry incorporated into the etching system it was able to control the etch and stop within the 15 nm W layer. In order to ensure electrical isolation and physical stability, the nanopillars were buried into 800 nm $SiO_2$ and planarized by ion beam milling with grazing incidence to expose the top of the pillar. The EBL was also used to define the HM line bottom electrode with a 6 µm length and width varying from 0.6 µm to 4.2 µm. Direct laser writing was used in the others lithographies in order to establish electrical contact with top and bottom electrodes.

After the nanofabrication, the devices were annealed at 300 ºC for 2 h, with an applied magnetic field of 1T along the same axis direction of the field used during the deposition in order to pin the synthetic antiferromagnetic layers. After the annealing the free layer of 1.4 nm $CoFe_{40}B_{20}$ exhibits in plane magnetic anisotropy[26].

## Data availability

The data that support the findings of this study are available from the corresponding author upon reasonable request.

## Acknowledgments

This work was supported in part by the Marie Sklodowska Curie Individual Fellowship (IF) for the SHADE project under the contract number of 897733 and in part by the European Union's Horizon 2020 FETOPEN Programme under project SpinAge, Grant ID 899559.

## Author contributions

HGH, HF and FM designed and performed the research, and wrote the manuscript together with TB, LB, and LB, PA, RF who fabricated the MTJ samples for testing and characterisation which was done by HGH, LB, PA, TB, and RF.

## Competing interests

The authors declare no competing interests.